\documentclass[12pt,preprint]{aastex}

\newcommand{\nh}{N_{H}}

\shorttitle{The evolutionary sequence of AGN}
\shortauthors{Page et al.}

\begin{document}


\title{The evolutionary sequence of active galactic nuclei 
and galaxy formation revealed}


\author{M. J. Page}
\affil{Mullard Space Science Laboratory, University College London,
Holmbury St Mary, Dorking, Surrey RH5 6NT, UK}
\email{mjp@mssl.ucl.ac.uk}

\author{J. A. Stevens and R. J. Ivison}
\affil{Astronomy Technology Centre, Royal Observatory, Blackford Hill, 
Edinburgh, EH9 3HJ, UK}
\email{jas@roe.ac.uk,rji@roe.ac.uk}

\and

\author{F. J. Carrera}
\affil{Instituto de F\'\i sica de Cantabria (Consejo Superior de
Investigaciones Cient\'\i ficas--Universidad de Cantabria), 39005
Santander, Spain}
\email{carreraf@ifca.unican.es}

\begin{abstract}
Today, almost every galaxy spheroid contains a massive black hole: a remnant
of, and testament to, a period in its evolution when it contained an active
galactic nucleus (AGN).  However, the sequence and timescales of the formation
of the black hole and surrounding spheroid of stars are completely unknown,
leaving a large gap in our knowledge of how the Universe attained its present
appearance.  Here we present submillimeter observations of matched samples of
X-ray absorbed and unabsorbed AGN which have luminosities and redshifts
characteristic of the sources responsible for most of the mass in present day
black holes. Strong submillimeter emission, an isotropic signature of copious
star formation, is found only in the X-ray absorbed sample, ruling out
orientation effects as the cause of the absorption.  The space density and
luminosity range of the X-ray absorbed AGN indicate that they are undergoing
the transition between a hidden growth phase and an unabsorbed AGN phase, and
implies that the X-ray absorbed period in the AGN's evolution coincides with
the formation of the galaxy spheroid.
\end{abstract}


\section{Introduction}

It is now known that the history of cosmic star formation and luminous AGN
activity track each other rather well, both showing a dramatic decline between
$z \sim 2$ and the present day \citep{franceschini99}.
Their legacy, the prevalence of massive black holes today \citep{yu02} and the
proportionality between black hole mass and the mass of their host galaxy
spheroids \citep{merritt01}, is most easily explained if the formation of the
two components was coeval, i.e. the black hole was built up by accretion of the
same gas that rapidly formed the stars of the spheroid.  During this formation
episode, the average bolometric luminosity of the stellar component is expected
to exceed that emitted by the AGN by a factor of a few \citep{page01sci}.
Under such circumstances,
a large fraction of stellar mass must have built up around the AGN that
dominate the accretion power of the Universe, because these were responsible
for the majority of present day black hole mass.  The bulk of the comoving
luminosity density was produced by objects with luminosities close to the break
luminosity $L_{*}$, on account of the AGN luminosity function at any redshift
being characterised by a broken power law \citep{page97,boyle00}.  While the
X-ray background is chiefly produced by AGN at $z<2$ 
\citep[e.g.][]{barger01}, the
luminosity density of AGN was at its peak between $z=1$ and $z=3$
\citep{page97,miyaji01}. Therefore coeval black hole / stellar bulge formation
would imply that a large fraction of the Universe's star formation took place
around AGN with $1<z<3$, and $\log(L) = \log(L_{*})\pm0.7$.

\begin{figure}
\epsscale{.60}
\plotone{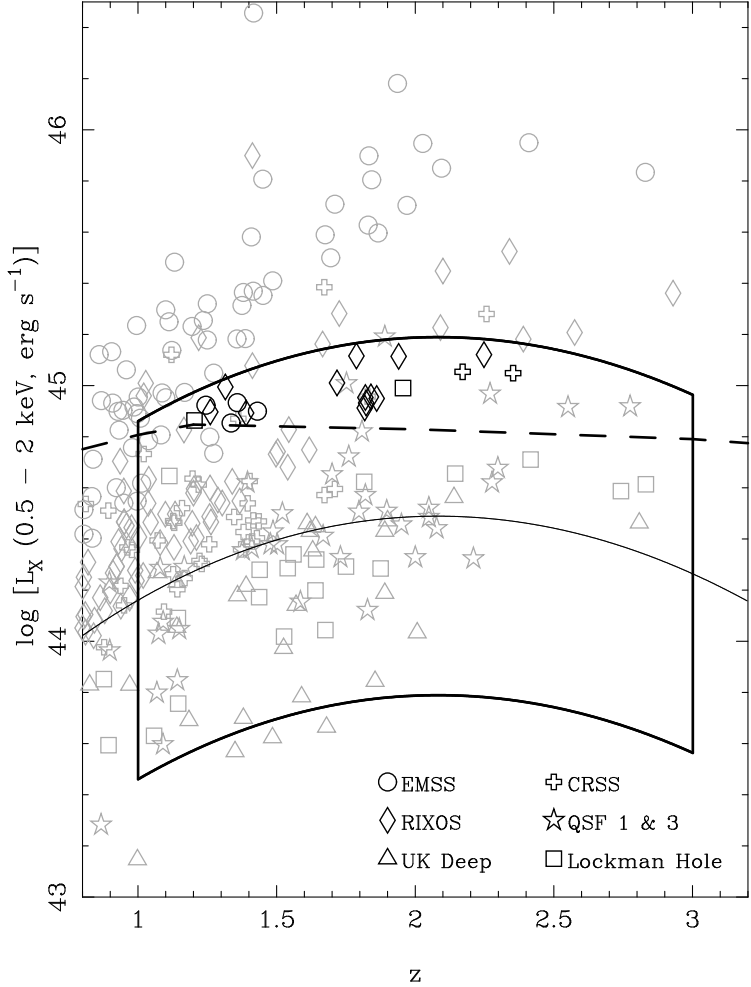}
\caption{Selection of the sample.  The datapoints show the redshifts and X-ray
luminosities of X-ray selected AGN from a number of surveys with different flux
limits and sky areas.
The thin solid line shows the
position of $\log(L_{*})$ as a function of redshift assuming the best fit model
luminosity function from \citet{page97}, 
 and the bold solid line encloses the region $(1<z<3, \log(L_{*})-0.7 < \log(L)
< \log(L_{*})+0.7)$, in which the majority of AGN comoving luminosity density
is produced.  The bold dashed line is the limit at which sources are expected
to have a flux of 3 mJy at 850$\mu$m if they have star-formation powered FIR
luminosities equal to their AGN bolometric luminosities, and assuming FIR
spectra similar to that of Mrk~231.  The AGN above the dashed line but within
the region enclosed by the bold line are the objects that can usefully be
probed with SCUBA {\em and} are relevant to the production of the Universe's
accretion luminosity. Sources selected for submillimeter observations are shown
in black rather than grey.  }
\label{fig:selection}
\end{figure}

Submillimeter observations have already shown that there is a significant
overlap between the bulge and black hole growth phases in X-ray absorbed QSOs
in this range of redshifts and luminosities \citep{page01sci}.  These enigmatic
objects are characterised by significant X-ray absorption (column densities of
$21.4< \log N_{H} < 22.5$) but little or no obscuration to their broad emission
lines and ultraviolet continua \citep{page01mn}.  However, sensitive
submillimeter observations have been performed for only eight such objects, and
hence the temporal overlap between black hole and bulge growth phases is
subject to considerable statistical uncertainty. More importantly, there were
no sensitive submillimeter observations of a comparable sample of unabsorbed
QSOs, which represent the majority of QSOs selected in soft X-ray surveys.
Therefore we followed up this
initial result with an experiment involving a larger sample of 20 unabsorbed 
objects. This
sample is matched in luminosity and redshift to the original sample; it
encompasses all of the crucial $1<z<3$, $\log(L) = \log(L_{*})\pm0.7$ region
which can viably be investigated with the Submillimeter Common User Bolometer
Array (SCUBA) at the James Clerk Maxwell Telescope (JCMT).  Figure
\ref{fig:selection} shows the region of redshift--luminosity space from which
the sample is drawn.  The sample was drawn from a combination of several
surveys: the Extended {\em Einstein} Medium Sensitivity Survey
\citep[EMSS,][]{stocke91}, the {\em Rosat} International X-ray Optical Survey
\citep[RIXOS,][]{mason00}, the UK {\em Rosat} Deep Survey \citep{mchardy98}, the
Cambridge-Cambridge {\em Rosat} 
Serendipity Survey \citep[CRSS,][]{boyle97}, the
QSF 1 \& QSF 3 fields \citep{griffiths95}, and the Lockman Hole
\citep{schmidt98}.  The wide range of flux limits covered by these surveys is
crucial in providing AGN over the required range of redshift-luminosity space.
In all practical respects this sample is comparable to the original sample,
except that it is composed of AGN that are unabsorbed in the X-ray band.
A Hubble constant $H_{0} = 70 {\rm km s^{-1} Mpc^{-1}}$ and density parameters
$\Omega_{\Lambda} = 0.7$ and $\Omega_{m} = 0.3$ are assumed throughout this Letter.

\section{Observations and data reduction}

Observations at 850$\mu$m were carried out at the JCMT between December 2002
and March 2003. The observations were carried out in photometry mode, in which
the source is placed on the central bolometer of the array and the secondary
mirror is jiggled in a 3 ${\times} 3$ pattern with 2 arcsec intervals with a 1
s integration at each position. The secondary mirror was chopped 45 arcsec in
azimuth at a frequency of 7.8 Hz and nodded between the source and reference
positions every 18 s.  Data were reduced using the standard STARLINK {\small
SURF} software. After compensating for the nod, the data were flatfielded and
corrected for atmospheric extinction. Each jiggle in turn was then corrected
for residual sky noise which is correlated across the SCUBA field of view and
often dominates the signal from faint sources. These sky offsets were taken to
be the median signal from all bolometers on the given array (excluding noisy
bolometers).

\section{Results and discussion}

\begin{figure}
\includegraphics[angle=270.0,scale=0.78]{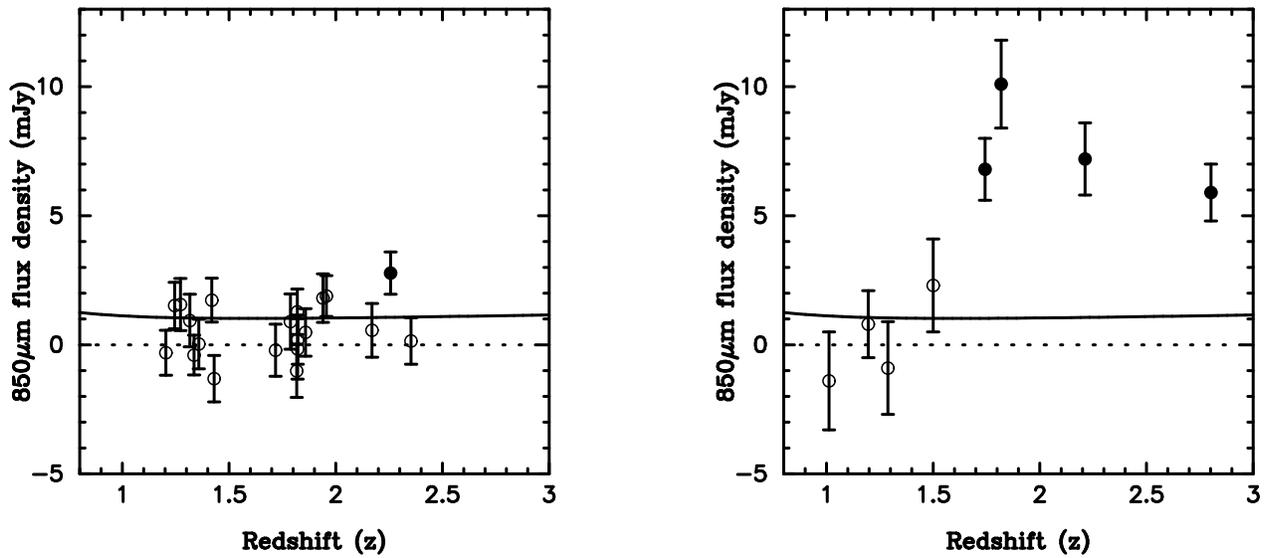}
\caption{Left panel: 850$\mu$m flux densities of the sample of 20 X-ray
unabsorbed QSOs. Right panel: 850$\mu$m flux densities of the matched sample of
X-ray absorbed QSOs,
adapted from \citet{page01sci}. In both panels, sources which are detected at
$>3\sigma$ confidence are plotted as filled circles, while non-detections are
shown as open circles.
The solid line on both panels shows the predicted 850$\mu$m flux density of the
nearby submillimeter luminous, X-ray absorbed QSO Mrk~231 if it were viewed at
redshift $z$.  }
\label{fig:hardflux}
\end{figure}

The measured 850$\mu$m flux densities for the X-ray unabsorbed QSOs are given
in Table \ref{tab:results}.  Only one source (RXJ141954) is detected with
$3\sigma$ confidence, and none of the sources are brighter than the
5\,$\sigma$/5\,mJy level.  This lack of detections is in stark contrast to the
sample of X-ray absorbed QSOs observed in 2001, of which 50\% are brighter than
5\,mJy (Figure \ref{fig:hardflux}). Far from simply improving the statistical
uncertainty inherent to our original sample, the results for the unabsorbed AGN
are incompatible with those for the absorbed sample.  The lack of 5\,mJy
detections in the unabsorbed sample implies that at 95\% confidence $<14\%$ of
unabsorbed QSOs are bright submillimeter sources \citep{gehrels86}, whereas the
detection rate of the X-ray absorbed sources implies that at least 19\% of
X-ray absorbed QSOs are bright submillimeter sources. 
 Our observations imply
that unabsorbed QSOs, which have been studied for 30 years, and of which we
know tens of thousands of examples, do not play a significant role in the
Universe's star formation history.  In contrast, the X-ray absorbed QSOs at
$z>1.5$ are embedded in the massive starbursts of still-forming galaxy
spheroids.

To see why this is such a suprising and consequential result, we must consider
the alternative models for the X-ray absorption in QSOs. By far the most
widespread and most paradigmatic model is the AGN `unified scheme'
\citep{antonucci93} in which the absorbing material is present in all objects
in the form of a toroidal structure. In this model, absorbed and unabsorbed AGN
are intrinsically identical, with the torus producing absorption in those
objects viewed through it.  This model is well established for the nearby,
lower luminosity Seyfert galaxies, although there are some notable 
discrepancies, \citep[e.g.][and references therein]{tran03}, 
and most X-ray background synthesis and AGN
population models are predicated on this concept
\citep{setti89,comastri95,gilli01}.  The more speculative alternative is that
X-ray absorbed and unabsorbed QSOs are intrinsically different objects. In
particular, it has been proposed that substantial absorption could be a
defining characteristic of the early phases of QSO evolution \citep{fabian99}.
In such models, the main, obscured growth phase of the QSO coincides with the
formation of the host galaxy spheroid, the completion of which coincides with
the luminous, unobscured phase of the QSO's evolution \citep{silk98,sanders88}.
In the unified scheme, any submillimeter emission powered by the nucleus will
be orientation independent because the obscuring torus is optically thin at
submillimeter wavelengths.  Submillimeter emission originating in star forming
regions will be similarly isotropic. Thus if the X-ray absorbed QSOs are
absorbed because of their orientation to our line of sight, their submillimeter
properties will be indistinguishable from those of X-ray unabsorbed QSOs.

Immediately then, the level of segregation we have found 
in the submillimeter properties of
the X-ray absorbed and unabsorbed sources allows us to rule out orientation as
the cause of the X-ray absorption. It therefore seems likely 
that the absorbed and unabsorbed
QSOs represent different stages in an evolutionary sequence, in which the
absorbed QSOs represent the earlier phase.  This hypothesis is confirmed by 
{\em Hubble Space Telescope} images which show that the majority of 
unabsorbed QSOs lie in
massive elliptical galaxies, at least for $0<z<1$ \citep{kukula01,dunlop03};
their regular, relaxed morphologies, imply that these are essentially mature,
finished galaxies. 
Direct measurements of the morphologies and 
stellar masses of the host galaxies of unabsorbed QSOs at $z>1$, while 
extremely difficult at present \citep[e.g.][]{kukula01,croom04},  
are also consistent with regular, relaxed morphologies.
The lack of submillimeter emission from our sample of
$1<z<3$ unabsorbed QSOs suggests that these objects also reside in relatively
quiescent, fully formed host galaxies. 
Therefore unabsorbed QSOs have already undergone a period of
prodigious star formation in which they would have shone as powerful
submillimeter sources. 
 That the X-ray absorbed QSOs are characterised by high
submillimeter luminosities implies that they are embedded in the dense
interstellar media of their forming host spheroids, and therefore that they are
still at this earlier evolutionary stage. Furthermore, the first submillimeter
images of their Mpc-scale environments indicate that they are located within
significant overdensities of ultraluminous starburst galaxies which will
rapidly evolve into rich clusters of galaxies typical of those that host X-ray
unabsorbed QSOs \citep{stevens04}.

We now argue that there is a particular time in the evolution of a typical 
QSO corresponding to the X-ray absorbed phase. It immediately precedes the
unabsorbed phase and lasts for only $\sim$15\% of the lifetime of the optical
QSO. 
First, their space density compared to unabsorbed QSOs
allows us to make a crude estimate of the relative durations of the two phases.
At a 0.5-2 keV flux of $10^{-14}$~erg~s$^{-1}$~cm$^{-2}$ the surface density of
X-ray sources with {\em Rosat} 
spectra similar to the absorbed QSOs is 19~deg$^{-2}$, of
which around half are broad-line objects \citep{page00,page01mn}.  The fluxes
must be corrected for absorption to allow a meaningful comparison with
unabsorbed QSOs; the average flux correction factor for the QSOs from the
hard-spectrum {\em Rosat} survey is 1.5 \citep{page01mn}.  Thus there are $\sim
9$ X-ray absorbed QSOs deg$^{-2}$ with an absorption corrected 0.5--2~keV flux
of $\ge 1.5\times 10^{-14}$~erg~s$^{-1}$~cm$^{-2}$.  This flux limit
corresponds to $L_{0.5-2}=4\times 10^{44}$~erg~s$^{-1}$ at $z=2$, i.e. close to
$L_{*}$ in the epoch when QSO activity peaked \citep{page97}.  For comparison,
there are $60-70$ unabsorbed QSOs at the same X-ray flux limit
\citep{page97,miyaji01}.  Assuming that absorbed and unabsorbed QSOs accrete
with the same Eddington ratio, this implies that the X-ray absorbed QSO phase
typically lasts only around 15\% as long as the unabsorbed phase (even less if
QSOs accrete with higher Eddington ratios while they are still absorbed).
Second, assuming that they accrete at 10\% efficiency, that they do not exceed
the Eddington limit, and that 3\% of the accretion power is emitted in the
0.5-2.0 keV band \citep{elvis94}, their X-ray luminosities imply that they
contain black holes of $>10^{8} M_{\odot}$. Thus they are already relatively
mature objects, and cannot grow by accretion much further without producing
black holes that are larger than those found in typical unabsorbed
QSOs. However, since their host spheroids are still forming, this brief phase
must precede the unabsorbed QSO with its complete, quiescent spheroid.  This
strongly suggests that the X-ray absorbed QSOs are observed during a
transitional phase which immediately preceeds the unabsorbed QSO, 
perhaps when the AGN has become powerful enough to eject
the remaining cold interstellar medium and thereby terminate the growth of the
spheroid \citep{fabian99,silk98}.

Before the X-ray absorbed phase, there is an initial phase that is so highly
obscured as to have been missed in all but the deepest X-ray surveys undertaken
so far. This earliest phase corresponds to the main growth period of the host
galaxy spheroid, and so should be luminous in the submillimeter. During this
phase the black hole must be accreting rapidly to achieve a mass of
$10^{8} M_{\odot}$, and so must reach X-ray luminosities in excess of $10^{44}
{\rm erg s^{-1}}$. If AGN in this initial phase were not X-ray
absorbed, then a significant fraction of unabsorbed QSOs would be luminous
submillimeter sources, but this is ruled out by our observations. If they were
absorbed to a similar degree as the X-ray absorbed QSOs, they would outnumber
this population substantially at the flux limits probed by the hard-spectrum
{\em Rosat} survey, and certainly would be found in large numbers just below
this flux limit in the deeper surveys that are now being carried out with {\em
XMM-Newton} and {\em Chandra}. This is not the case \citep{mainieri02,page03},
and hence the earlier evolutionary phases must be characterised by considerably
higher levels of X-ray absorption ($\nh > 10^{23} {\rm cm}^{-2}$).  Since the
great majority of optically selected QSOs show little or no X-ray absorption
\citep{yuan98}, the earlier phases must also be heavily obscured in the
restframe optical/UV.  Consequently, AGN in the earliest stages of their
evolution will be difficult to detect in any waveband, except for the
far-IR/submillimeter where they will be strong sources. At present, the 
small number of
optically faint, hard X-ray detected submillimeter sources reported by
\citet{ivison02} and \citet{alexander03} are probably the best candidates
for AGN in this initial phase because they are more heavily obscured than 
our {\em Rosat}-selected objects, although we can anticipate that
future high-throughput X-ray observatories could reveal the buried AGN 
components in many more submillimeter sources.  In contrast,
the X-ray absorbed QSOs found in the hard-spectrum {\em Rosat} survey may be
caught at the earliest evolutionary stage readily accessible to conventional
X-ray and optical surveys; flourishing and youthful, they may only just be
emerging from the obscuring gaseous cocoons of their birth.

\acknowledgments

The James Clerk Maxwell Telescope is operated on behalf of the Particle Physics
and Astronomy Research Council of the United Kingdom, the Netherlands
Organisation for Scientific Research and the National Research Council of
Canada. FJC acknowledge financial support by
the Spanish Ministerio de Ciencia y Tecnolog\'\i{}a, under grant ESP2003-00812.

\clearpage



\newpage

\clearpage

\begin{deluxetable}{lccr}
\tabletypesize{\normalsize}
\tablecaption{Characteristics of the unabsorbed sample and their observed
submillimeter emission. $L_{X}$ is the 0.5-2 keV luminosity in erg~s$^{-1}$;
$S_{850}$ is the 850$\mu$m flux density in mJy. Bold text indicates the 
source detected at $3\sigma$ confidence at 850$\mu$m.}
\tablewidth{0pt}
\tablehead{
\colhead{Source} & \colhead{$z$} & \colhead{log $L_{X}$}& \colhead{$S_{850}$}
}
\startdata
RX J105316.8+573552  &  1.204 & 44.9 &  $-0.3\pm 0.9$ \\
MS 0537.4-2843       &  1.245 & 44.9 &  $ 1.5\pm 0.9$ \\
RX J125456.9+564941  &  1.261 & 44.9 &  $ 1.6\pm 1.0$ \\
RX J125639.5+472410  &  1.315 & 45.0 &  $ 0.9\pm 1.0$ \\
MS 1633.1+2643       &  1.336 & 44.9 &  $-0.4\pm 0.8$ \\
MS 1703.5+6052       &  1.358 & 44.9 &  $ 0.0\pm 1.0$ \\
RX J071858.8+712432  &  1.390 & 44.9 &  $ 1.7\pm 0.9$ \\
MS 0824.2+0327       &  1.431 & 44.9 &  $-1.3\pm 0.9$ \\
RX J100926.3+533424  &  1.718 & 45.0 &  $-0.2\pm 1.0$ \\
RX J105649.6+493412  &  1.788 & 45.1 &  $ 0.9\pm 1.1$ \\
RX J111918.6+211339  &  1.818 & 44.9 &  $-0.2\pm 1.2$ \\
RX J180534.0+694733  &  1.820 & 45.0 &  $-1.0\pm 1.0$ \\
RX J104718.6+541919  &  1.823 & 44.9 &  $ 0.2\pm 1.0$ \\
RX J232443.6+231537  &  1.840 & 45.0 &  $ 1.3\pm 0.9$ \\
RX J113655.3+295131  &  1.861 & 44.9 &  $ 0.5\pm 0.9$ \\
RX J112106.0+133825  &  1.940 & 45.1 &  $ 1.8\pm 0.9$ \\
RX J105331.8+572454  &  1.956 & 45.0 &  $ 1.9\pm 0.8$ \\
CRSS J1428.3+4231    &  2.171 & 45.1 &  $ 0.6\pm 1.0$ \\
{\bf RX J141954.3+543014}  &  {\bf 2.248} & {\bf 45.1} &  ${\bf  2.8\pm 0.8}$\\
CRSS J1415.1+1140    &  2.353 & 45.0 &  $ 0.2\pm 0.9$ \\
\enddata
\label{tab:results}
\end{deluxetable}






\end{document}